\documentclass[review,number,sort&compress]{elsarticle}
\usepackage{subcaption}
\journal{Elsevier}

\begin{document}

\begin{frontmatter}

\title{A New Method for Evaluating the Effectiveness of Plastic Packaging Against Radon Penetration}

\author{Yue Meng\corref{cor1}\fnref{fn1}}
\fntext[fn1]{Now at Shanghai Key Laboratory for Particle Physics and Cosmology, Institute of Nuclear and Particle Physics (INPAC) and School of Physics and Astronomy, Shanghai Jiao Tong University, Shanghai 200240, China}
\ead{mengyue@sjtu.edu.cn}
\author{Jerry Busenitz \corref{cor2}}
\author{Andreas Piepke \corref{cor3}}
\address{Department of Physics and Astronomy, University of Alabama, Tuscaloosa, AL 35487, US}
\cortext[cor1]{Corresponding author}
\begin{abstract}
Deposition of $^{222}Rn$ daughters onto detector materials pose a risk to ultra-low background experiments. To mitigate this risk, a common approach is to enclose detector components in sealed plastic bags made of films known to be effective barriers against radon. We describe a new method to evaluate radon barriers which is unique in that (a) it gauges not only the intrinsic resistance to radon penetration of a plastic film but also the integrity of bags fabricated from the film and sealed following some protocol, and (b) it employs gamma spectroscopy rather than alpha spectroscopy.  We report the results of applying this method to sealed bags fabricated from  polypropylene, Nylon, Mylar, metallized Mylar, FEP, and PFA.   Evaluation of the fluoropolymers FEP and PFA as radon barriers are the first such measurements.
\end{abstract}

\end{frontmatter}
\section{Introduction}
For low background dark matter or neutrino experiments, $^{222}Rn$ daughter ``plate-out" onto the surface of detector components is a potentially dangerous source of experiment backgrounds.  The main process is neutron production via the $(\alpha,n)$ process induced by $^{210}Po$ decay. A common prevention method 
is to pack materials and parts inside sealed bags which are known to be effective barriers against radon diffusion. We apply a Pylon radon source (Pylon Model RN-1025), an air-tight purge box, bags fabricated from different materials, a steel can with a press-fit lid, and a high purity germanium detector to measure radon penetration through sealed bags. 
In addition to providing a quantitative measure of the effectiveness of a packaging method, the measurement results may be interpreted in terms of the radon diffusion constant and radon solubility for the bag material; we are providing such interpretation in Section~\ref{sec:permeability}.    

Many studies of films as radon barriers have been carried out; see, for example, \cite{doi:10.1063/1.4818090, doi:10.1093/rpd/ncr482, malki2011, Mamedov_2011} and citations therein. What is unique to this method is that it employs gamma spectroscopy and evaluates not only the film but also sealed bags fabricated from these films.
\section{Experimental Setup and Procedure}
\begin{figure}[!th]
\includegraphics[width=13cm]{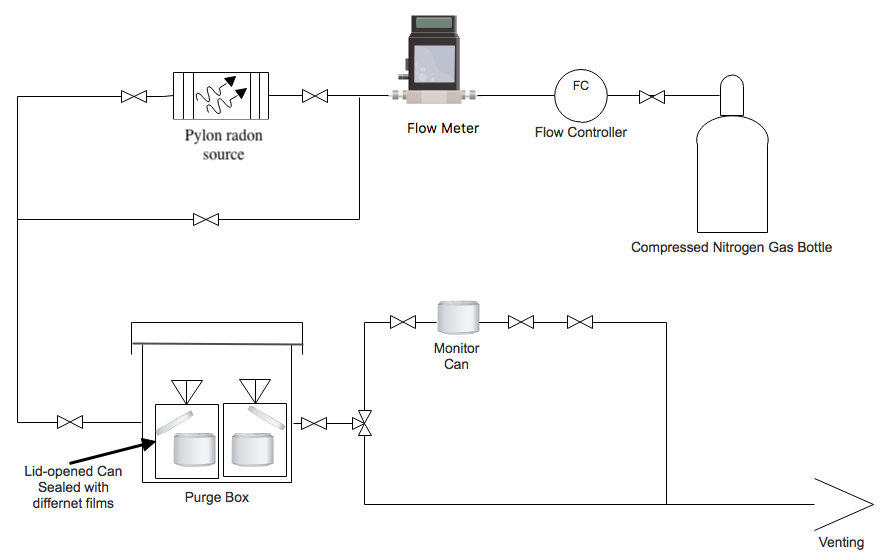}\\
\caption{The radon purge system.}
\label{fig:container}
\end{figure}

Radon penetration measurements through different sealed bags were conducted at room temperature. The purge system (Figure \ref{fig:container}) consists of a carrier gas bottle, a flow controller, a flow meter, a 
radon source, an air-tight purge box and a ``monitor can''.
The monitor can contains a sample of the radon loaded gas used to determine film penetration. At the end of the radon exposure it is counted and serves as normalization for the ``sample can'', described below. 
An open steel can, equipped with a press-fit lid, is sealed (along with the lid) inside a test bag procured commercially. This test bag is made of the film to be studied. After placing the steel can and lid inside the bag, the open end is closed by making three adjacent seals with a hand-operated impulse heat sealer.  This ``sample can''-bag assembly is, in turn, placed inside an air-tight purge box where it is exposed to a radon loaded gas atmosphere for some period of time. Operation in gas purge mode assures a constant radon concentration. The lateral dimensions of the sealed bag were measured to determine its area $A$. The volume $V$ of the bag interior was estimated, for some bags, by immersing representative samples in water and measuring the displacement.  Radon was transferred to the purge box by means of nitrogen carrier gas, flowing through a Pylon (Model RN-2015) radon source, containing 246 $kBq$ of $^{226}Ra$. The radon concentration of the purge gas is given by the calibration certificate of the source. 
Multiple samples were exposed to this $^{222}Rn$--rich atmosphere sufficiently long for radon concentrations to reach steady--state.  The radon purge was then terminated by closing off and detaching the monitor can and quickly flushing the purge box with nitrogen gas. The bag - sample can combinations were then removed from the purge box. To seal the gas, contained inside the bag, into the sample can, the lid was pressed in, before opening and discarding the bag.
The radon activity contained inside the, now sealed, ``sample cans'' was determined using HP Ge-detectors. This counting was performed in form a so-called time series, allowing a fit to the exponential decay, to determine the $^{214}Pb$ and $^{214}Bi$ activities at the reference time, defined as the end of the radon purge.
\section{Principle}
\label{sec:principle}
The measured radon concentration inside the bag, relative to that outside the bag during exposure to the radon--rich atmosphere, provides a measure of the effectiveness of the sealed bag as a radon barrier.  Assuming that the bag has been sealed properly, the effectiveness depends mainly on the diffusion of radon through the film, which is described in terms of the radon diffusion constant and the solubility of radon in the film material.   In this section, we  derive the relationships used to interpret the ratio of measured concentrations in terms of the diffusion constant and solubility.  See Reference~\cite{FERNANDEZ2004167} for an alternate derivation with details.  

The following assumptions are made. First, the time over which the bag is exposed to the radon--rich environment is sufficiently long for steady--state conditions to be achieved. Once the diffusion constant is determined, the time required to reach steady--state can be estimated and compared to the actual exposure time to check this assumption. Second, the film may be treated as a planar barrier, a good approximation if the characteristic radius of curvature of the bag surface is large compared to the film thickness. For our measurements, the typical film thickness is on the level of a few mils ($\sim$100~microns) and the typical radius of curvature is on the level of a few cm. Third, the concentration of radon inside the bag is instantaneously uniform throughout the volume. This is effectively achieved at room temperature where the lineal dimension--about 10 cm--of the bag divided by root mean square speed of the radon atoms is short compared to the diffusion time through the film. Last, radon emanation from the bag or its contents is negligible.
Radon diffusion through a planar barrier, where $x$ is the distance into the barrier measured from the outside surface, obeys the diffusion equation, with the additional term $\lambda\cdot C$, to account for the decay of the diffusing substance:
\begin{equation}
{{\partial C}\over{\partial t}} = D\cdot {{\partial ^2 C}\over{\partial x^2}} - \lambda \cdot C,
\end{equation}
where $C$ is the radon number concentration (atoms per unit volume), $D$ is the diffusion constant and $\lambda$ is the radon decay constant. We emphasize that $C$ is the radon number concentration within the barrier material itself.\\
At steady state 
\begin{equation}
0= D\cdot {{\partial ^2 C}\over{\partial x^2}} - \lambda \cdot C
\end{equation}%
Let $C _0 \equiv C(0) $ be the radon number concentration at the outside edge of the film  and $C _1  \equiv C(d) $ be the radon number concentration at the inside edge of the film, where we have denoted the film thickness by $d$.  The concentration of radon atoms outside the bag will not necessarily be the same as $C _0$, owing to the fact that the solubility of radon in the bag material may be different than it is in the outside medium.  For the same reason, $C_1$ will not necessarily be the same as the radon number concentration inside the bag. We therefore introduce the external and internal gas-space radon number  concentrations, $C _0 ^\prime$ and $C _1 ^\prime$, respectively. $C _0 ^\prime$ depends on the properties of the radon purge system and $C _1 ^\prime$ is measured. Finally, we link the boundary values of the radon number concentrations in the bag film with the radon number concentrations in the internal and external media by 
\begin{equation}
{{C _0}\over{C _0 ^\prime}} = {{C _1}\over{C _1 ^\prime}} = S
\end{equation}
where we have assumed that the gases on the inside and outside of the bag are identical. $S$ is the solubility of radon in the bag film relative to the gas.  For our measurements, the outer medium is nitrogen gas and the inner medium initially air. One expects that the inner medium becomes dominantly nitrogen gas in steady state.\\
As already mentioned, $C_0 ^{\prime}$ is inferred from the properties of the radon purge system (radon source calibration and carrier gas flow rate) and $C_1 ^{\prime}$ is measured.  At steady--state, the number of radon atoms diffusing into the interior of the bag equals the decay rate inside the bag. $\phi _C$, the flux of radon atoms at depth $x$ in the film, is evaluated according using Fick's law: 
\begin{equation}
\phi _C = -D\cdot {{dC}\over{dx}}.
\end{equation}
Inside the sealed bag of total surface area $A$ and volume $V$, the concentration $C_1 ^{\prime}$ obeys the equation
\begin{equation}
{{dC_1 ^{\prime}}\over{dt}} = \left(\phi_{C_1}\cdot {{A}\over{V}} -\lambda\cdot C_1 ^\prime \right),
\end{equation}
where $\phi_{C_1}$ denotes the radon flux at the inner surface of the bag. The change in radon concentration inside the bag receives a contribution from diffusion into the bag and another one from Rn-decay. It is assumed that the bag itself is not a source of radon.\\
Once steady state has been achieved,
\begin{equation}
\phi_{C_1} = \lambda\cdot {{V}\over{A}}\cdot C_1 ^{\prime} 
\end{equation}
Combining the above relationships, we obtain
\begin{equation}
\alpha\cdot \beta = {{S\cdot A}\over{V}}\left[{{1 - \beta\cdot \cosh (\alpha\cdot d)}\over{\sinh (\alpha\cdot d)}}\right]
\label{eqn:pervssolubility}
\end{equation}
where
\begin{equation}
\beta \equiv {{C _1 ^\prime}\over{C_0 ^\prime}}
\end{equation}
\begin{equation}
\alpha \equiv \sqrt[]{\frac{\lambda}{D}}
\end{equation}
This transcendental equation, which agrees with Reference \cite{FERNANDEZ2004167} but uses different notation, can be solved for the diffusion constant if the ratio of concentrations and solubility are known.  Alternately, if the 
ratio of concentrations and radon diffusion constant are known, one may determine the solubility.
In order to determine both the radon diffusion constant and solubility, one must measure the concentrations as a function of time, not just at steady--state, as reported here.   Therefore, our measurements determine the radon diffusion constant as a function of the solubility.
\section{Measurement of Relative Radon Penetration through Packaging Materials}
\label{sec:relative}

\begin{figure}[!h]
\centering
\begin{subfigure}{.5\textwidth}
  \centering
  \includegraphics[width=6.3cm]{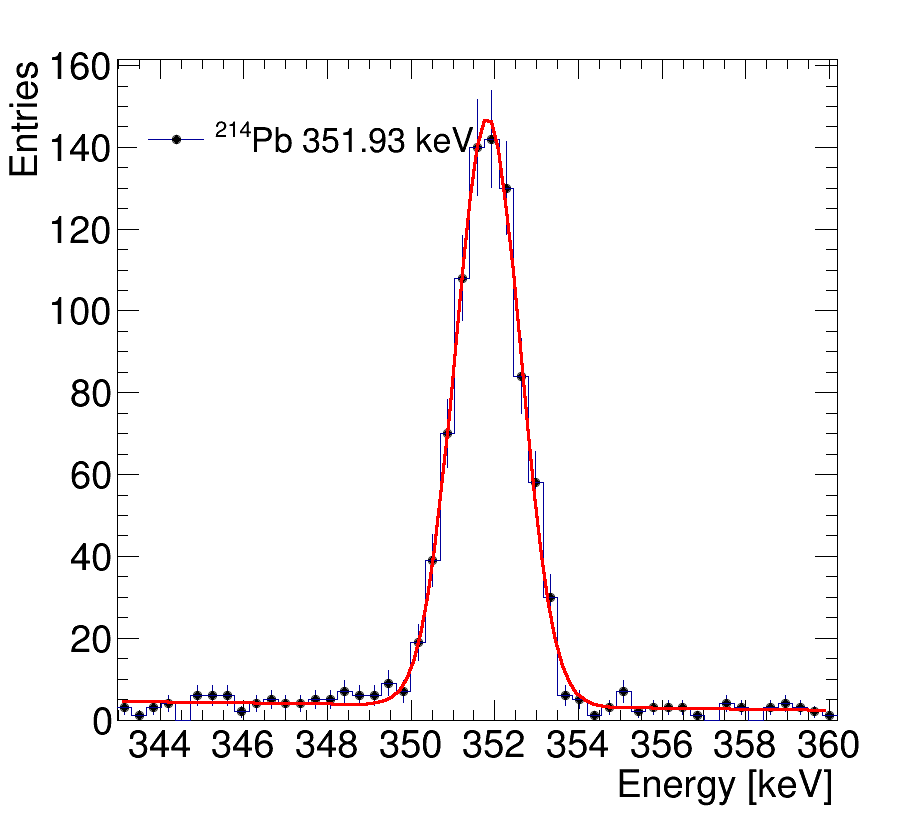}
  \label{fig:352}
\end{subfigure}%
\begin{subfigure}{.5\textwidth}
  \centering
  \includegraphics[width=6.3cm]{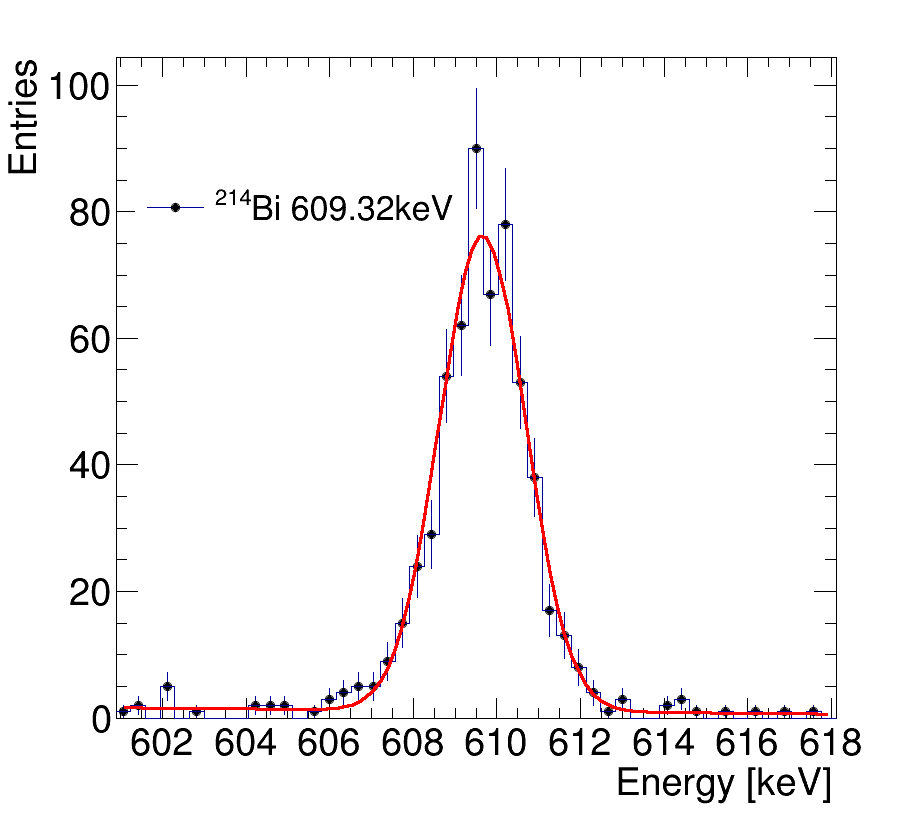}
  \label{fig:609}
\end{subfigure}
\caption{Left: 352 $keV$ $\gamma$-peak of $^{214}Pb$. Right: 609 $keV$ $\gamma$-peak of $^{214}Bi$. Black dots are data and the red lines show the fits.}
\label{fig:spectra}
\end{figure}

%
Since 4--5 samples were exposed in each radon purge run, each sample was counted, in turn, with the counting time interval for each sample chosen based on the expected activity.  For each daughter, only the most prominent gamma line, 352 $keV$ for $^{214}Pb$ and 609 $keV$ for $^{214}Bi$ were used.  Example fits to $^{214}Pb$ and $^{214}Bi$ $\gamma$-peaks (Gaussian plus linear background) are shown in Figure \ref{fig:spectra}.



\begin{figure}[!th]
\centering
\begin{subfigure}{.5\textwidth}
  \centering
  \includegraphics[width=6cm]{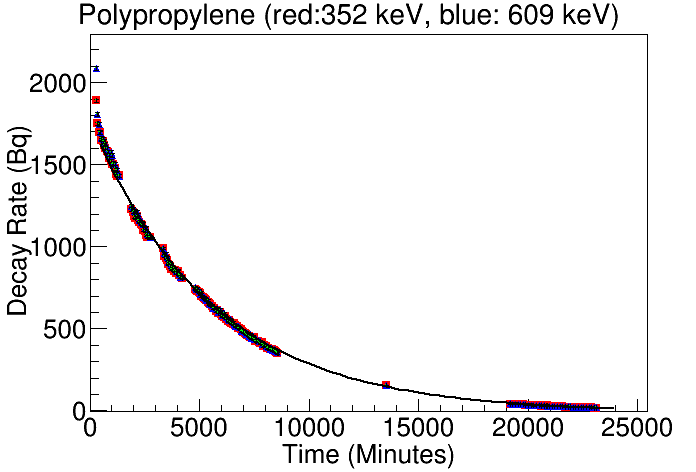}
  \caption{Polypropylene packing material}
  \label{fig:sub1}
\end{subfigure}%
\begin{subfigure}{.5\textwidth}
  \centering
  \includegraphics[width=6cm]{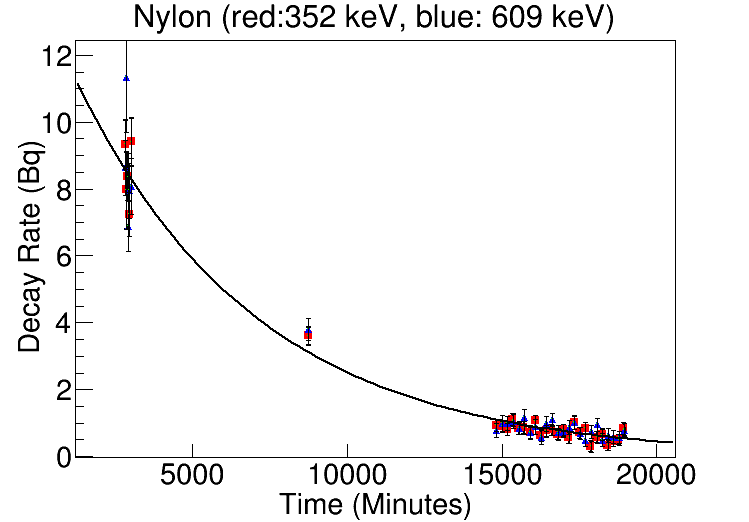}
  \caption{Nylon packaging material}
  \label{fig:sub2}
\end{subfigure}
\caption{$^{214}Pb$ and $^{214}Bi$ decay rates vs time. The red dots show the time dependence of the 352 $keV$ $\gamma$-peak rates, resulting from $^{214}Pb$ decay. The blue dots that of the 609 $keV$ $\gamma$-peak of $^{214}Bi$. Both are fitted (solid black line) with an exponential plus constant background.}
\label{fig:decayvstime}
\end{figure}

The decay rate of $^{214}Pb$ and $^{214}Bi$, contained in the monitor and sample cans, can be inferred from the counting data, live-times, branching ratios and detection efficiencies \cite{Tsang:2019apx}. This determines the $^{222}Rn$ activity contained in the cans.
Example decay curves for sample cans sealed inside polypropylene and Nylon pouches
are shown in Figure \ref{fig:decayvstime}. An exponential function plus a constant background 
was used to fit the sequential decay rates in Figure \ref{fig:decayvstime}.   Excluding the first few hours after end of purge allows for secular equilibrium to be established. 
 
The effective half--life measured for each sample fell in the range 2.6--2.9 days, to be compared to the $^{222}Rn$ half--life of 3.8 days.  
We interpret this difference as evidence for a slow leak from the cans, due to the press fit lid not being radon--tight. Under this leak hypothesis the gas loss corresponds to a half life of about 11 days.

Four groups of measurements were performed. 
Each series of measurements was devoted to a different packaging material. Polypropylene bags served as reference sample in each group. 
The results of this first series of measurements, in which the monitor can had not yet been implemented, are reported as a ratio $R$ of the activity determined for the can sealed inside the bag material under study, divided by the activity of a can exposed in parallel but contained in a polypropylene bag. Polypropylene was chosen to serve as the reference material as it is known to be quite permeable to radon \cite{doi:10.1093/rpd/ncr482}. The $R$-values obtained this way are shown in Table~\ref{tab:baginfo} together with other relevant parameters.

\begin{table}[htbp]
\begin{center}
\scalebox{0.88}{
\begin{tabular}{c|c|c|c|c}
\hline
Bag material & \begin{tabular}{@{}c@{}}Bag volume \\$[L]$\end{tabular} & \begin{tabular}{@{}c@{}}Bag area \\$[cm^2]$ \end{tabular}  & \begin{tabular}{@{}c@{}} Film thickness \\$[\mu m]$\end{tabular}& $R$ \\ \hline 
\multicolumn{5}{c}{Group 1} \\ \hline
Polypropylene      & 0.7   & 697  & 101.6 & 1 \\ \hline
Nylon                    & 0.4   & 407   & 50.8  & 0.0104 $\pm$ 0.0003 \\ \hline
Nylon                    & 0.4   & 407   & 50.8  & 0.0076 $\pm$ 0.0003 \\ \hline
Metallized Mylar-Type 1  & 0.5   & 503  & 63.5  & 0.0002 $\pm$ 0.0001 \\ \hline 
Metallized Mylar-Type 1  & 0.5   & 503  & 63.5  & 0.0005 $\pm$ 0.0001\\ \hline 
\multicolumn{5}{c}{Group 2} \\ \hline
Polypropylene      & 0.8   & 929  & 101.6 & 1 \\ \hline
Metallized Mylar-Type 2      & 0.5	&  542  &	101.6   & 0.0009 $\pm$ 0.0003 \\ \hline
Metallized Mylar-Type 3      & 0.5 &  542  & 109.2  &  0.0012 $\pm$ 0.0003 \\ \hline 
\multicolumn{5}{c}{Group 3} \\ \hline
Polypropylene      & 0.7   & 668  & 101.6 & 1 \\ \hline
PFA & 0.4   & 387  & 50.8   & 1.17 $\pm$ 0.01 \\ \hline
PFA & 0.4   & 387  &  50.8   &  1.15 $\pm$ 0.01\\ \hline
FEP                      & 0.4 	& 439  & 127.0   & 0.114 $\pm$ 0.001\\ \hline
FEP                     & 0.4 	& 411  & 127.0   & 0.110 $\pm$ 0.002\\ \hline
Transparent Mylar                      & 0.4   & 439  & 76.2  & 0.0006 $\pm$ 0.0002\\ \hline 
Transparent Mylar                      & 0.4   & 411  & 76.2  & 0.0009 $\pm$ 0.0003\\  \hline 
\multicolumn{5}{c}{Group 4} \\ \hline
Polypropylene      & 0.4   & 397  & 101.6 & 1 \\ \hline
Transparent Mylar      & 0.4   & 397  & 76.2  & 0.0001 $\pm$ 0.0004\\ \hline
Nylon*      & 0.4   & 397  & 50.8  & 0.99 $\pm$ 0.01\\ \hline
Metallized Mylar-Type 1      & 0.4   & 397  & 63.5  & 0.0011 $\pm$ 0.0002\\ \hline
Metallized Mylar-Type 2      & 0.4   & 397  &	101.6   & 0.0019 $\pm$ 0.0008 \\ \hline
Metallized Mylar-Type 3      & 0.4   & 397  & 109.2	 & 0.0027 $\pm$ 0.0003 \\ \hline 
\end{tabular}
}
\end{center}
\caption{Packaging properties and results for groups of samples.  The left column gives the packaging material. (For FEP and PFA, bag fabrication and sealing was performed by American Durafilm Co., Holliston, MA, USA.) The bag area and volume are estimated from lateral dimensions and in some cases water displacement measurements.   Column 4 lists the film thickness. $R$ is the ratio of the radon concentration, determined for the sample can contained in the material of interest, divided by the concentration measured for a sample can packaged in polypropylene film. The errors given with the $R$-values are statistical only.
 *See the following section.}
\label{tab:baginfo}
\end{table}%
\section{Bag Integrity Testing}
\begin{figure}[!th]
\begin{center}
\includegraphics[width=7cm]{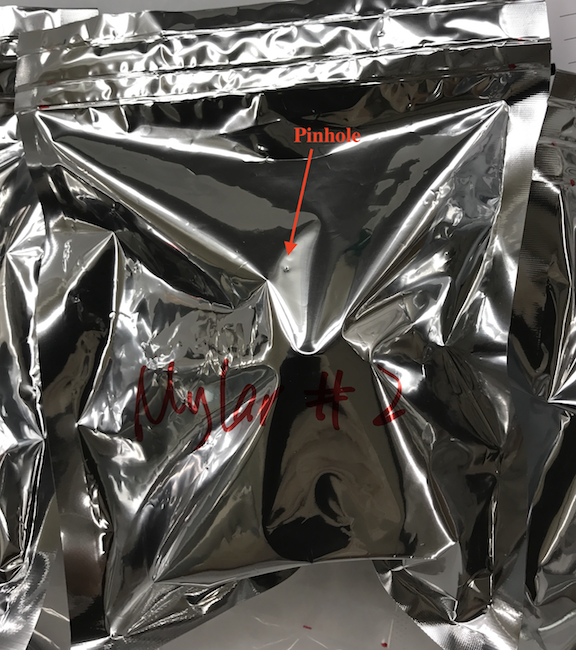}\\
\caption{Meltallized Mylar bag with a pinhole, indicated by the  red arrow. The diameter of the pinhole is less than 0.5 $mm$.}
\label{fig:pinhole}
\end{center}
\end{figure}

The data presented in Table~\ref{tab:baginfo} shows one relatively high $R$-value for a Nylon bag tested in Group 4. This reading seems to contradict the Nylon results obtained in Group 1. It should be noted that all Nylon bags were made from the same material stock. The high reading further seems to run counter previous permeability measurements made by others~\cite{WOJCIK2000158,WOJCIK2004355}.
We take this result as evidence that this particular bag was not properly sealed.  We carried out a 3--week radon exposure of three metallized Mylar bags and a polypropylene reference bag. Two of the metallized Mylar bags had pinholes (Figure \ref{fig:pinhole}) punched into them to demonstrate the method presented here is capable of identifying them.    
All three metallized Mylar bags were sealed shut.  
  The counting results and bag dimensions are listed in Table \ref{tab:pinholetale}. This data clearly identifies the punctured Mylar bags as such. 
  This data shows that the method described here is capable of identifying even small breaches in radon enclosure films.
\begin{table}[!htbp]
\begin{center}
\scalebox{0.8}{
\begin{tabular}{c|c|c|c|c}
\hline
Bag material & \begin{tabular}{@{}c@{}}Bag volume \\$[L]$\end{tabular} & \begin{tabular}{@{}c@{}}Bag area \\$[cm^2]$ \end{tabular}  & \begin{tabular}{@{}c@{}} Film thickness \\$[\mu m]$\end{tabular}& \begin{tabular}{@{}c@{}}Radon concentration \\$[Bq/can]$\end{tabular} \\ \hline 
Polypropylene     & 0.7 & 697  & 101.6 & 567.9 $\pm$ 1.2\\ \hline
Metallized Mylar (bag 1) & 0.4  & 542  & 63.5 & 1.4 $\pm$ 0.6\\ \hline 
\begin{tabular}{@{}c@{}}Metallized Mylar (bag 2) \\ with pinhole \end{tabular}  & 0.4   & 542  & 63.5  & 509.8 $\pm$ 11.9\\ \hline
\begin{tabular}{@{}c@{}}Metallized Mylar (bag 3) \\ with pinhole \end{tabular}  & 0.4   & 542  & 63.5  & 642.3 $\pm$ 3.9 \\ \hline 
\end{tabular}
}
\end{center}
\caption{Radon concentrations and bag geometries for the ``pinhole test'' measurements.}
\label{tab:pinholetale}
\end{table}%
\\
\section{Ratio of Concentrations Relative to Monitor Can}
\label{sec:absolute}
As described in Section \ref{sec:principle}, the method discussed here can be used to estimate the diffusion constant if the solubility of radon in the bag material is known. This requires, however, that the ratio of the radon concentrations inside and outside of the bag are known.
In order to interpret our radon permeation results in terms of terms of diffusion constants, we implemented a ``monitor can'' for absolute normalization, as mentioned earlier. With it we sample the radon-loaded gas directly. We denote the ratio of concentrations determined for a particular sample relative to the monitor can by $R'$.
A polypropylene can was used in this test as well. This measurement of the ratio of the polypropylene activity to the monitor can activity, denoted $R_{pp}$ allows one to re-normalize the $R$-data reported in Table~\ref{tab:baginfo} by the simple scaling relation: 
\begin{equation}
R ^{\prime} = R _{PP}\cdot R
\end{equation}
To determine $R _{PP}$, 5 different polypropylene-sealed sample cans were analyzed.
The data resulting from these measurements is shown in Table \ref{tab:ppper}. This data shows good consistency and reproducibility. An average variance-weighted correction factor of $R_{PP} = 0.856 \pm 0.004$ was obtained this way.

\begin{table}[!thbp]
\begin{center}
\scalebox{0.87}{
\begin{tabular}{c|c|c|c|c|c}
\hline
Bag material & \begin{tabular}{@{}c@{}}Bag volume \\$[L]$\end{tabular} & \begin{tabular}{@{}c@{}}Bag area \\$[cm^2]$ \end{tabular}  & \begin{tabular}{@{}c@{}} Film thickness \\$[\mu m]$\end{tabular}& \begin{tabular}{@{}c@{}}Concentration \\$[Bq/can]$\end{tabular} & $R_{PP}$ \\ \hline 
Monitor can        &        &         &       & 516.5 $\pm$ 2.3 & 1 \\ \hline
Polypropylene      & 0.9   & 813   & 101.6 & 479.1 $\pm$ 3.6 & 0.88 $\pm$ 0.01\\ \hline
Polypropylene      & 0.9   & 813   & 101.6 & 465.0 $\pm$ 2.7 & 0.85 $\pm$ 0.01\\ \hline
Polypropylene      & 0.9   & 813   & 101.6 & 454.9 $\pm$ 1.9 & 0.84 $\pm$ 0.01\\ \hline
Polypropylene      & 0.9   & 813   & 101.6 & 466.9 $\pm$ 1.4 & 0.86 $\pm$ 0.01\\ \hline
Polypropylene      & 0.9   & 813   & 101.6 & 464.1 $\pm$ 1.3 & 0.85 $\pm$ 0.01\\ \hline 
\end{tabular}
}
\end{center}
\caption{Determination of the scaling factor $R_{PP}$ (defined in the text) by means of repeated measurements. Uncertainties are statistical only.}
\label{tab:ppper}
\end{table}%


Table \ref{tab:ratio} shows the $R_{PP}$-corrected activity ratios for the films studied here. In case more than one measurement were available, the variance-weighted average was taken and renormalized by the correction factor $R_{PP}$.


\begin{table}[!tbhp]
\begin{center}
\begin{tabular}{c|c}
\hline
Bag material & \begin{tabular}{@{}c@{}}$R^{\prime}$\end{tabular}  \\ \hline
Polypropylene & $0.856\pm 0.004$  \\ \hline
Transparent Mylar & $0.0005\pm 0.0002$\\ \hline
Metallized Mylar-Type 1 & $0.0004\pm 0.0001$\\ \hline
Metallized Mylar-Type 2 & $0.0009\pm 0.0002$ \\ \hline
Metallized Mylar-Type 3  & $0.0017\pm 0.0002$\\ \hline
PFA & $0.99\pm0.01$  \\ \hline
FEP & $0.097\pm0.001$  \\ \hline
Nylon & $0.0077\pm 0.002$  \\ \hline
\end{tabular}
\end{center}
\caption{Properly corrected measured ratios of radon concentrations inside and outside of bags made from the materials listed here.}
\label{tab:ratio}
\end{table}%
\section{Determination of Diffusion Constants for Comparison with Previous Measurements}
\label{sec:permeability}
Diffusion constants for some of the materials investigated here have been previously measured.
In order to demonstrate the validity of
our method, we used our data to estimate the diffusion constant. This allows for a consistency check with previous measurements in the cases of polypropylene, Nylon, and transparent Mylar.

\begin{figure}[!thbp]
\begin{center}
\includegraphics[width=12cm]{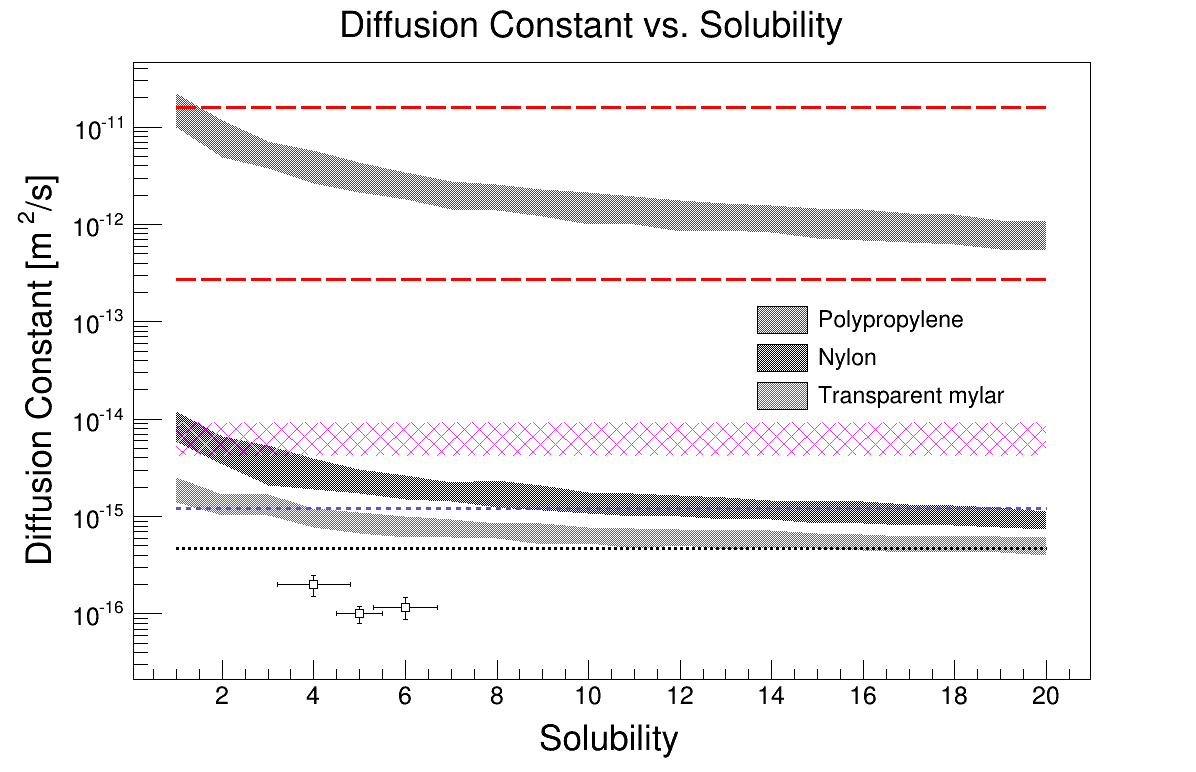}\\
\caption{Diffusion constants versus solubility of polypropylene (red band), Nylon (black band) and transparent Mylar (blue band), estimated from the data presented in this paper.  Also shown (by points, lines, and hatched bands) are results from previous measurements.   See the text for details. }
\label{fig:pervssol}
\end{center}
\end{figure}

Using Equation \ref{eqn:pervssolubility}, 
we estimate the diffusion constant for each material as a function of the unknown solubility
from our activity ratio, film thickness, bag area and bag enclosed volume data.
The diffusion constant was calculated varying the solubility over a range of 1--20.  Estimated uncertainties on volume, surface area, and concentration ratio allow us to construct the $\pm1\; \sigma$ confidence bands shown in Figure~\ref{fig:pervssol} (red: polypropylene, black: Nylon, blue: transparent Mylar).

The red dashed lines in Figure~\ref{fig:pervssol} indicate the minimum and maximum values for the diffusion constant for
polypropylene as reported in Reference~\cite{doi:10.1093/rpd/ncr482}.  
The dashed blue line shows the upper limit on the diffusion constant of transparent Mylar, as published in \cite{doi:10.1063/1.4818090}.
There are several previous measurements for Nylon, including ones in which the diffusion constant and solubility were simultaneously measured. 
The dotted black line denotes the value of the diffusion constant reported in \cite{doi:10.1063/1.4818090}. The results of References~\cite{WOJCIK2000158} and \cite{WOJCIK2004355} for dry Nylon are plotted as open--square points. Shown by the pink hatched band is
the measurement of the dry Nylon diffusion constant reported by Reference~\cite{supernemo_2015}.

The diffusion constants estimated from the data presented in this paper are, within a reasonable range of solubility, consistent with previous measurements. Nylon seems to be an exception.  
A possible reason for this deviation could be
uncontrolled experimental parameters.
It was reported in reference~\cite{WOJCIK2000158} that the diffusion constant of Nylon depends on the water content of the material. Uncontrolled environmental humidity may, thus, serve as an explanation. Note, however, that Reference \cite{supernemo_2015} did not observe a dependence on relative humidity.  It is further not clear by how much the Nylon diffusion constant depends on the manufacturing details which we don't control.
We therefore conclude that such measurements allow one to {\it estimate} the barrier effectiveness for various film materials while {\it precise measurements} require tight material specifications and controls. The latter was beyond the scope of the study presented here.

\section{Conclusion}
We describe a relatively simple method for evaluating the effectiveness of sealed bags as radon barriers.  
We show that this method is very sensitive for detecting even small breaches in the barrier.
Results are
reported for polypropylene, FEP, PFA, Nylon, transparent Mylar, and several types of metallized Mylar.  

\section{Acknowledgements}
This work was motivated by the need of the LZ collaboration to effectively seal delicate detector components from environmental radon and the resulting plate-out of its daughters.
We thank our LZ colleagues for  
encouragement and stimulating discussions.  We are grateful to Devin Radloff for his contributions to the initial setup, measurements, and data analysis. This research was supported in part by the U.S. Department of Energy under DOE Grant DE-SC0012447.  
\bibliography{mybibfile}
\end{document}